\documentclass[aip,jap,reprint]{revtex4-1}

\usepackage{amsmath}
\usepackage{epsfig,graphicx}
\usepackage[dvips,ps2pdf]{hyperref}

\newcommand{\be}{\begin{equation}}
\newcommand{\ee}{\end{equation}}

\begin{document}

\title{Laser-induced charging of microfabricated ion traps}

\author{Shannon X. Wang}   \email[]{sxwang@mit.edu}
\author{Guang Hao Low}
\author{Nathan S. Lachenmyer}
\author{Yufei Ge}
\author{Peter F. Herskind}
\author{\\Isaac L. Chuang}
\affiliation{Center for Ultracold Atoms, Research Laboratory of Electronics and Department of Physics, Massachusetts Institute of Technology, Cambridge, Massachusetts, 02139, USA}

\date{\today}

\begin{abstract}
\noindent 
Electrical charging of metal surfaces due to photoelectric generation of carriers is of concern in trapped ion quantum computation systems, due to the high sensitivity of the ions' motional quantum states to deformation of the trapping potential.  The charging induced by typical laser frequencies involved in doppler cooling and quantum control is studied here, with microfabricated surface-electrode traps made of aluminum, copper, and gold, operated at 6 K with a single Sr$^+$ ion trapped 100 $\mu$m above the trap surface. The lasers used are at 370, 405, 460, and 674 nm, and the typical photon flux at the trap is 10$^{14}$ photons/cm$^2$/sec. Charging is detected by monitoring the ion's micromotion signal, which is related to the number of charges created on the trap. A wavelength and material dependence of the charging behavior is observed: lasers at lower wavelengths cause more charging, and aluminum exhibits more charging than copper or gold. We describe the charging dynamic based on a rate-equation approach.

\end{abstract}

\maketitle

\section{Introduction}  

Microfabricated ion traps are promising candidates for realizing large-scale quantum computers\cite{Home:09}. Recent efforts have concentrated on development of multi-zone surface-electrode ion traps with small trap sizes\cite{Leibrandt:09, Amini:10}, so that traditional microfabrication techniques can be employed. The typical ion-to-metal distance in these traps is on the order of 10-100 $\mu$m, small enough that the trapped ions are sensitive to surface effects such as electric-field noise (causing anomalous heating) and localized charging of the trap electrodes or substrate. While anomalous heating of ions trapped in microfabricated traps has been studied extensively both by theory and experiment \cite{Turchette:00, Dubessy:09, Labaziewicz:08b}, laser-induced charging has only seen a few systematic experiments recently. So far, laser-induced charging has been studied on the glass subtrate of planar gold traps \cite{Debatin:08}, on copper traps including insulators brought close to the trap surface \cite{Harlander:10}, and aluminum traps \cite{Allcock:11}. Several unknown issues, including material dependence and the role of oxide layers on the metal, remain. For example, no such studies of charging have been done on aluminum traps with varying oxide layers, or comparisons made between different electrode materials with the same experimental setup. 

The lasers used for any typical ion trap experiment span a wide range of wavelengths. In a microfabricated trap, they are much closer to the trap surface, and as traps become smaller in size, it is increasingly difficult to avoid scatter caused by lasers illuminating the trap. In some experiments, lasers are deliberately shone onto the trap for the purpose of micromotion compensation\cite{Allcock:10}. This could be expected to cause buildup of electrical charges on the trap surface due to the photoelectric effect. The typical shortest wavelengths needed for ion traps range from 194~nm for Hg$^+$ to 493~nm for Ba$^+$, corresponding to 6.4-2.5~eV. Typical work functions for metals used for ion traps such as Au, Ag, Al, Cu, etc are $\sim$4~eV or higher, but may change due to surface effects such as the presence of an oxide layer. 

The choice of material for ion traps is an important consideration. Gold has been a popular choice due to its chemical inertness, and it has a high work function of greater than 5~eV, but is incompatible with traditional CMOS fabrication. Consequently there has been some interest in using aluminum\cite{Leibrandt:09} or copper for microfabricated ion traps, which can take advantage of sophisticated CMOS fabrication techniques. Pure aluminum has a high work function at 4.2~eV and is expected not to release electrons when illuminated with light at 405~nm for Sr$^+$. However, aluminum is also known to quickly form a native oxide layer, Al$_2$O$_3$, which may lower the work function and thus make it susceptible to blue light. Such effects have been observed in previous studies of the photoelectrochemical effects of blue light on aluminum and other materials \cite{Semov:69, DiQuarto:91}. Local charges formed on the Al$_2$O$_3$ may not dissipate, changing the trapping potential and leading to excess micromotion \cite{Berkeland:98}, which can affect the stability of the trap. 

\begin{figure}
\includegraphics[width=3.375in]{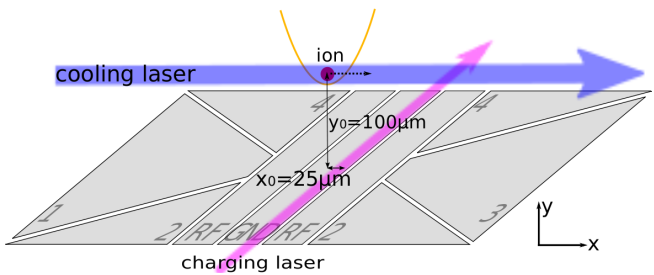}
\caption{\label{Fig:3ddiag} Diagram of charging experiment setup (not to scale). The 4 DC compensation electrodes are labelled 1, 2, 3, 4. The charging laser is displaced along the $x$ axis as shown by $x_0 = 25$~$\mu$m, such that the ion's displacement has a non-zero projection (dotted line) along the direction of the cooling laser. The axes' origin is taken to be the point along the charging laser's waist nearest to the ion (endpoint of the right arrow of $x_0$).}
\end{figure}

In this work, we study the charging behavior of aluminum, copper, and gold microfabricated traps when illuminated with lasers at 674, 460, 405, and 370~nm. All traps are operated in a cryogenic system at 6~K. Charging is measured by observing the micromotion amplitude of a single trapped $^{88}$Sr$^+$ ion and relating it to ion displacement. In the aluminum traps, we find a wavelength dependence of the charging behavior: the laser at 405~nm charges the trap noticeably on timescales of minutes, whereas minimal charging is observed with 460~nm and 674~nm lasers over the same timescales. Copper traps exhibit charging at all wavelengths. No charging is observed at any of these wavelengths for gold traps, but some is observed at 370~nm. A schematic of our charging laser and trap geometry is shown in Fig.~\ref{Fig:3ddiag}.

We describe these experiments beginning in section \ref{sec:model}, which covers the physical model, the charging dynamics using rate equations, and the measurement method. Section \ref{sec:expt} covers trap fabrication and the experimental setup and measurement. Section \ref{sec:results} describes the results and presents numerical estimates for the relative charging of different materials and wavelengths.

\section{Model} 
\label{sec:model}

We postulate a basic model of the charging process as the photoelectric effect on a metal modified by a thin-film oxide layer, similar to the approach taken in Ref.\cite{Harlander:10}. Electron-hole pairs are created near the metal-oxide interface with an initial rate that is proportional to the power of the incident light. As electrons accumulate in the oxide layer, the charging rate decreases due to screening. At the same time, the electrons diffuse at a rate set by the material properties of the oxide layer. We assume that the dissipation of holes in the metal is much faster than the rate of electron diffusion and screening, due to the higher conductivity of the metal.

Based on the simple picture of the photoelectric effect modified by oxides, one would expect that light of lower wavelength and materials with oxide layers or lower conductivity would charge more.

\begin{figure}
\includegraphics[width=3in]{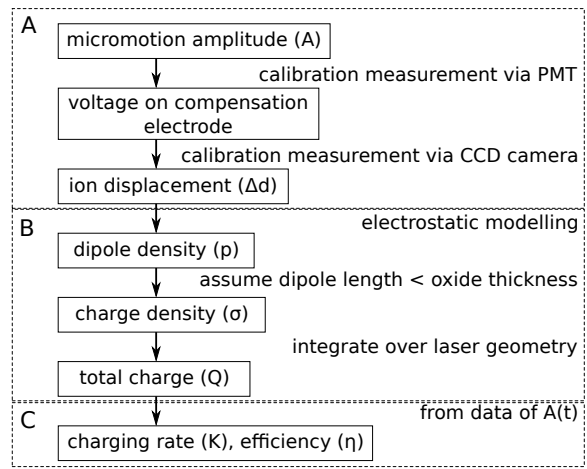}
\caption{\label{Fig:block}Block diagram illustrating the conversion between the measured quantity, micromotion amplitude $A(t)$, and the desired quantities, (A) ion displacement $\Delta d$ and (B) total charge Q. }
\end{figure}

Here we describe the relations between the measured quantities and physical parameters in the model (see Fig.~\ref{Fig:block}) and the rate-equation model used to fit the time evolution of the charging behavior. Section \ref{sec:modelMM} defines the micromotion amplitude, relating it to the ion displacement and electric field. Section \ref{sec:iondisp} describes the conversion from ion displacement and electric field to a quantitative estimate of charges on the trap. Section \ref{sec:modelRate} describes the charging dyamic using a charge accumulation rate, a dissipation rate and screening rate as parameters which leads to a rate equation for fitting to the measured micrmotion amplitude vs time.

\subsection{Micromotion and ion displacement}  
\label{sec:modelMM}

The dynamics of a trapped ion is described in Ref\cite{Berkeland:98} and the relevant parts are summarized here. The motion of a single trapped ion in a Paul trap with a quadratic pseudopotential is characterized by a low-frequency ``secular'' oscillation and an oscillation called ``micromotion'' at the frequency of the applied rf field. For a surface-electrode trap, the trapping potential is slightly modified\cite{House:08, Wesenberg:08}, but the nature of the motion (with two characteristic frequencies) is the same. The intrinsic micromotion which occurs when the secular motion carries the ion through the nodal point of the rf field (rf null) is small and will not be of concern in this work. We focus on the ``excess'' micromotion discussed next. 

Assuming the ion is initially located in the rf null such that no micromotion is present, any additional charges generate an electric field which displaces the potential minimum point such that the ion is no longer located in the rf null. With an ion displacement of $\Delta d$, the micromotion amplitude is $\frac{q_i}{2}|\Delta d|$ along the direction of displacement, where $q_i$ is the Mathieu $q$ parameter along the same axis. This excess micromotion cannot be significantly reduced by Doppler cooling because it is driven by the rf field\cite{Berkeland:98}. Experiments generally seek to minimize micromotion due to its effect on spectral properties of the ion \cite{Berkeland:98}, but here we take advantage of the well-defined temporal behavior of micromotion to discern small displacements in the ion position. This technique is closely related to the Doppler velocimetry technique that has recently been used for ultra-sensitive force detection in Penning traps \cite{Biercuk:10}.

\begin{figure}
\includegraphics[width=3.375in]{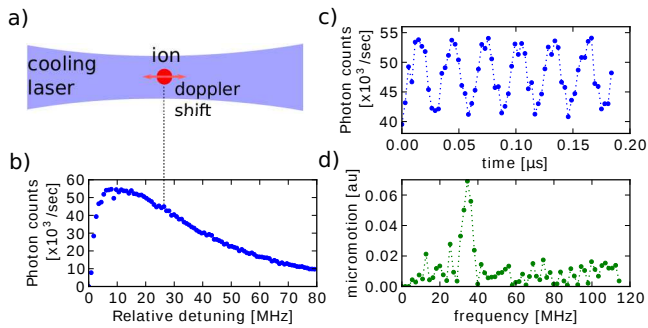}
\caption{\label{Fig:samplePMT} Measurement of micromotion signal. (a) Diagram of an ion executing micromotion along the direction of a Doppler cooling laser beam. (b) Typical scan of the fluorescence curve. The dotted line between (a) and (b) indicates the ion's scattering rate in the absence of micromotion. (c) The oscillating fluorescence signal due to Doppler shift from the micromotion. Photon counts are normalized. (d) Fast Fourier transform of the fluorescence signal normalized to total fluorescence. The maximum value gives the micromotion amplitude. }
\end{figure}

Micromotion of the ion is measured using the fluorescence detection method\cite{Berkeland:98} (see Fig.~\ref{Fig:samplePMT}). A photomultiplier tube (PMT) detects fluorescence of the ion, and single photon arrival times are binned to 3~ns bins. This is fast enough to capture the modulation of the fluorescence due to the Doppler shift at near the rf drive frequency of the trap, which is typically between 34-37~MHz. The amplitude of these oscillations, $A(t)$, gives a measure of the amplitude of the micromotion along the propagation direction of the cooling laser, and is obtained by performing a fast Fourier transform of the PMT signal. This observed amplitude is proportional to the ion displacement $\Delta d$, as verified by two calibration measurements described in detail in Section \ref{sec:calib}. The micromotion amplitude is observed to vary linearly with the applied voltage on one of the compensation electrodes (electrode 1, see Fig.~\ref{Fig:3ddiag}), and the ion displacement also varies linearly with this voltage. 

\subsection{Ion displacement and charge distribution} 
\label{sec:iondisp}

In the approximation of a harmonic potential, the ion displacement $\Delta d$ can be related to the electric field $\mathbf{E}$ at the ion location generated by the laser-induced charges as $\Delta d = e \mathbf{E}\cdot \hat{x}/m \omega^2$, where $e$ is the ion's charge, $m$ is the ion's mass, and $\omega$ is the secular frequency along the direction of the ion's displacement\cite{Berkeland:98}. For simplicity we only consider the ion displacement along $\hat{x}$, the radial axis parallel to the trap surface (axis $x$ in Fig.~\ref{Fig:3ddiag}), so that all the analysis can be done in the one-dimensional model.

The laser-induced charges are located above a conducting surface and thus should be considered to be dipoles, due to the image charge induced in the conductor\cite{Harlander:10}. The size of the dipole $r_d$ in the expression for the dipole moment, $q r_d$,  is unknown and thus the number of dipoles created by the laser cannot be easily determined with the techniques described here; however, a rough estimate of the order-of-magnitude of the charge generation rate can be obtained by bounding the dipole size by twice the thickness of the oxide layer. For aluminum, the thickness is taken to be 3-5~nm from the literature\cite{Campbell:99}. The growth of oxide on copper is not self-limiting as in aluminum and thus its thickness is difficult to estimate; it is assumed that gold has no native oxide layer.

The spatial distribution of the laser-induced charge is taken to be an area of dipoles as follows. The laser intensity distribution on the trap at grazing incidence can be approximated as a line with constant intensity along the trap axis and gaussian distributed intensity profile along the axis perpendicular to it, with waist (radius) $\omega_0$. The Rayleigh range of all charging lasers is longer than the length of the trap, so intensity variations along the axial direction can be ignored. We approximate the distribution of charges created on the trap as a Gaussian along the radial direction and constant along the axial direction, directly proportional to the laser intensity. Let $p$ be the dipole moment density, which is related to the charge density $\sigma$ as $p = \sigma r_d$ where $r_d$ is the size of the dipole. The potential due to such an infinite gaussian line of dipoles is given by:
\be V_{\rm line}(x_0,y_0) = \frac{1}{\sqrt{2\pi  }\sigma }\int _{-\infty }^{\infty }e^{-\frac{2x^2}{\omega_0 ^2}}\frac{p y_0}{2 \pi  ((x-x_0)^2+y_0^2) \epsilon _0} dx
 \ee
where $x_0$ and $y_0$ are the horizontal and vertical displacement of the charges from the ion respectively (see Fig.~\ref{Fig:3ddiag}), and $\epsilon_0$ is the vacuum permittivity constant. 

To summarize, from the ion displacement $\Delta d$ we obtain the electric field and thus the potential created by the laser-induced charges at the ion's location. By assuming a spatial distribution, the potential can be converted to dipole and charge density.

\subsection{Charge accumulation \& dissipation} 
\label{sec:modelRate}

\begin{figure}
\includegraphics[width=3.3in]{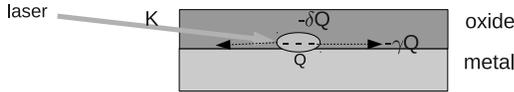}
\caption{\label{Fig:diag} Illustration of rate constants in the model of charge accumulation and dissipation. $K$ is the rate of electron creation, $-\delta Q$ is the modification to the charging rate due to screening, and $-\gamma Q$ is the rate of discharging through the oxide.}
\end{figure}

Let $Q(t)$ be the amount of charge present in the oxide layer generated by a laser incident on the trap as a function of time, with $Q(0)=0$. The charging rate is modeled by two processes. Let $K$ be the (constant) rate of charge accumulation due to the incident laser. The presence of existing electrons modifies the charging rate over time due to screening, represented by a rate $-\delta Q$. Discharging through the oxide can be modeled by $ -\gamma Q$ where $\gamma$ is a constant set by material properties \cite{Debatin:08}. Solving the rate equation $\dot{Q} = K-\delta Q - \gamma Q$ with $Q(0) = 0$ gives $Q(t)=\frac{K}{\delta + \gamma} (1-e^{-\delta t - \gamma t})$. The time constant for this charging/discharging process is then $\tau = 1 / (\delta+ \gamma)$. At 6~K the conductivity of insulators is expected to be lower than at room temperature, leading to a longer time constant of discharging. Fig.~\ref{Fig:diag} illustrates the model and rates.

Let $A(t)$ be the measured micromotion amplitude as a function of time and $A_\infty$ be the limiting value of $A$ as $t\rightarrow \infty$. The relation between the measured micromotion amplitude $A(t)$ and charge $Q(t)$, therefore $K$ and $A_\infty$, is obtained as follows: micromotion amplitude $\rightarrow$ voltage on compensation electrode $\rightarrow$ ion displacement and electric field $\rightarrow$ dipole density $\rightarrow$ charge density $\rightarrow$ total charge (see Fig.~\ref{Fig:block}). The measured micromotion amplitude vs time can thus be written as:
\be \label{eqn:At} A(t) = A_\infty (1-e^{-\frac{t}{\tau}})  \ee
where $A_\infty$ is the saturated micromotion amplitude (as t$\rightarrow\infty$) and is proportional to the term $\frac{K}{\delta+\gamma}$. We use this phenomenological model to fit the experimental data of micromotion vs time and extract the values of $\Delta d$ at saturation (corresponding to $A_\infty$) and the time constant $\tau$. Finally we estimate the initial charging rate $K$ and the charging efficiency $\eta$, the latter defined as the number of charges created per photon at $t=0$.

\section{Experiment} 
\label{sec:expt}

The fabrication of the surface-electrode traps used in this work follows the standard optical lithography procedures described in Section \ref{sec:fab}. The experimental setup and measurement method are described in Section \ref{sec:setup}. Section \ref{sec:calib} describes the calibration measurements to convert the observed micromotion amplitude to ion displacement and electric field.

\subsection{Trap fabrication}  
\label{sec:fab}
\begin{figure}
\includegraphics[width=3.36in]{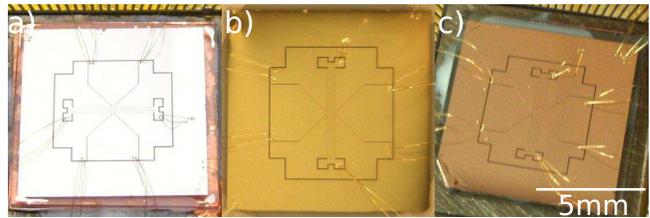}
\caption{\label{Fig:traps} Photographs of (a) aluminum, (b) gold, and (c) copper traps. }
\end{figure}

We fabricate 1 aluminum, 1 copper, and 2 gold traps for the charging tests described here. In addition we fabricate 3 aluminum traps with additional deposited layers of oxide in thicknesses of 5, 10, and 20~nm. The traps are of a 5-rod surface-electrode design\cite{Labaziewicz:08}. All traps are fabricated with optical lithography on 0.5~mm-thick quartz substrates. The aluminum trap with no additional oxide layer is made by first evaporating 1~$\mu$m of aluminum on the substrate at a rate of 0.45~nm/s. After lithography using NR9-3000P photoresist, the trap is patterned with wet chemical etch using Transene aluminum etchant type A. No attempt is made to modify the native aluminum oxide formed via contact with air between fabrication and testing. Aluminum traps with extra layers of deposited oxide are fabricated using the lift-off process. After lithography on photoresist, 400~nm of aluminum is evaporated at a rate of 0.33~nm/s, followed by 5, 10, or 20~nm of Al$_2$O$_3$ at a rate of 0.11~nm/s. After fabrication, the traps are coated with a protective layer of photoresist. Copper and gold traps with electrode thicknesses of 400~$\mu$m are fabricated using a very similar lift-off process, except that a 10~nm initial layer of Ti is needed for adhesion during evaporation. Photos of the traps are shown in Fig.~\ref{Fig:traps}.

It is well-known that surfaces exposed to ordinary laboratory environments absorb a few monolayers of hydrocarbon contaminants within a few hours \cite{Degen:09, Kuehn:06}. In our experiment, no attempt was made to clean the trap surfaces in situ, so it may be argued that surface contaminants will play a role in the charging effects that we observe. In addition, the exposed dielectrics between electrodes have also been suspected to contribute significantly to charging. To minimize the effects of varying surface preparation, the protective layer of photoresist on all traps is removed only immediately before packaging and installing. The process for packaging and installing in vacuum takes between 12-16 hours.

\subsection{Experimental setup \& methods} 
\label{sec:setup}
The trap is cooled and operated in a 4~K bath cryostat\cite{Antohi:09}. Loading is done via photoionization of a thermal vapor. Typical distance between the ion and the trap surface is $y_0 = 100$~$\mu$m. Doppler cooling is performed on the 422~nm S$_{1/2}$--P$_{1/2}$ transition\cite{Labaziewicz:07}. All lasers used for trapping propagate parallel to the trap surface. The typical axial secular frequency is 800~kHz and radial secular frequencies are 1-1.5~MHz. A CCD camera is used to image the ion, in conjunction with fluorescence detection by a PMT. 

Lasers at 674, 460, 405, and 370~nm are used for the charging measurements. They propagate along the axial direction of the trap. For the measurements, they are brought to grazing incidence on the trap as confirmed by observing their scatter on the trap surface using the CCD camera. The 370, 405, and 460~nm lasers have a beam diameter of 100~$\mu$m and 100~$\mu$W of power. The 674~nm laser has a beam diameter of 34~$\mu$m and power of 200~$\mu$W. Based on the geometry of the experiment we estimate the grazing incidence angle to be no more than 1 degree. From this we can calculate the peak photon flux to be $\sim 10^{14}$ cm$^{-2}$ s$^{-1}$. The lasers are incident on the trap with a horizontal offset of $x_0 = 25(5)$~$\mu$m from directly below the ion, such that there is a discernible displacement of the ion along the radial axis, $x$. A schematic of the trap and laser beams is shown in Fig.~\ref{Fig:3ddiag}.

In the absence of deliberate charging by aligning the laser to graze the trap surface, the ion's micromotion signal and compensation voltages are observed to be stable for a long time, on the order of a day. Before each charging measurement, the micromotion of the ion is minimized by applying voltages on the four DC compensation electrodes. Typically the observable micromotion along the radial or axial direction of the trap is sensitive to a 0.01~V change in the compensation voltages. For the measurements described here, we focus only on the micromotion caused by the radial displacement of the ion, parallel to the trap surface.

\subsection{Calibration}  
\label{sec:calib}
\begin{figure}
\includegraphics[width=3.3in]{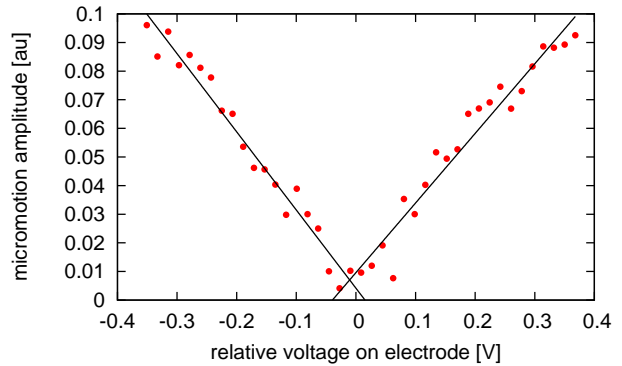}
\caption{\label{Fig:mmvscomp} Micromotion amplitude vs voltage (relative to optimal compensation) on a compensation electrode, used to convert the measured micromotion amplitude to ion displacement. Lines are linear fits.}
\end{figure}

The observed micromotion amplitude is converted to the displacement of the ion and electric-field changes at the ion location via calibration measurements and modelling of the trap potential. We calibrate the micromotion amplitude to the voltage applied to one of the compensation electrodes, 1. The voltage on electrode 1 is scanned and the resulting changes in micromotion is measured as shown in Fig.~\ref{Fig:mmvscomp}. Linear fits to this data give the conversion between micromotion amplitude and voltage on the compensation electrode 1, $c_1=0.24(1)$~[au]/V. The ion displacement as a function of voltage is also measured by applying a voltage on the electrode and measuring the ion displacement with the CCD camera. The resolution of the imaging optics is insufficient for measuring the ion displacement directly during charging measurements, so a larger voltage must be applied to obtain this calibration. The ion is displaced $c_2=0.75(3)$~$\mu$m/V. From electrostatic modelling of the trap geometry, for an ion height of 100~$\mu$m, the electric-field sensitivity is 50(2)~V/m for every 1~V applied to the electrode. From these calibration, the fitting parameter $A_\infty$ can be converted to ion displacement: $\Delta d = (1/c_1)c_2A_\infty$.

\section{Results} 
\label{sec:results}

In Section \ref{sec:resultsA} we describe the materal and wavelength dependence of the observed charging behavior. Section \ref{sec:resultsB} describes the measurement to determine the sign of charge, as well as estimates for the initial charging rate $K$ and efficiency $\eta$. 

\subsection{Material \& wavelength dependence}  
\label{sec:resultsA}

\begin{figure}
\includegraphics[width=3.3in]{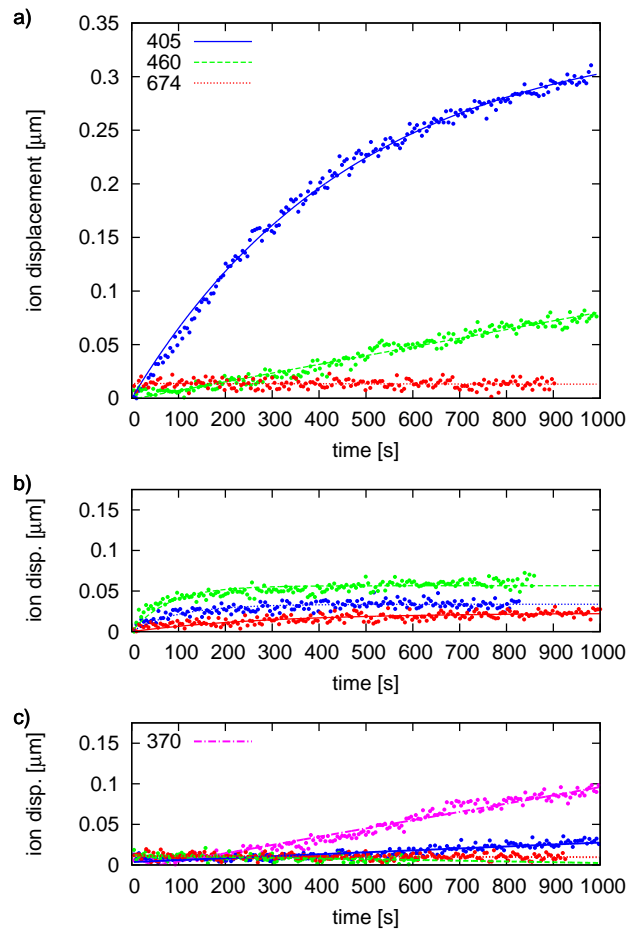}
\caption{\label{Fig:Chdata} Typical plot of ion displacement over time in (a) aluminum, (b) copper, and (c) gold traps showing charging dynamic for all wavelengths: 405~nm (blue, top, solid), 460~nm (green, middle, dashed), 674~nm (red, bottom, dotted), and 370~nm (magenta, dash-dot). Data is smoothed over 5 second intervals. Plots for 405~nm for the aluminum trap and all wavelengths for copper are fit to Eq.~\ref{eqn:At}.}
\end{figure}

We tested one trap each of copper, gold, and one each of aluminum with different thicknesses of oxide layers: native oxide, 10, and 20~nm. Aluminum traps, both with and without the additional oxide layer, exhibit charging behavior when the 405~nm laser is incident on the trap. Fig.~\ref{Fig:Chdata} shows measured evolutions of micromotion amplitude over time as the laser is turned on at $t=0$ for each of the trap materials. The two parameters that describe the micromotion amplitude over time, $A_\infty$ (or $\Delta d$) and $\tau$, are obtained by fitting Eq.~\ref{eqn:At} to the data. For the aluminum traps and 405~nm, typical time constants are 400-800~s and the saturated micromotion amplitude corresponds to an ion displacement of $\Delta d = 0.34(3)$~$\mu$m at the end of the measurement time, or an electric field at the ion location of $\sim$20~V/m. No significant variation of the charging rate or time constant as a function of oxide thickness is observed. The ion displacements measured here are about an order of magnitude smaller than those reported in Harlander et al\cite{Harlander:10}, likely due to the much smaller trap size and differences in the laser/trap geometry. The electric field is slightly smaller than that observed in Allcock et al\cite{Allcock:11}, where again a different trap geometry and laser wavelengths are used. 

In the copper trap, the most pronounced charging effect is observed with the 460~nm laser. Less charging is observed with the 405 and 674~nm lasers. The reversed wavelength dependence of charging in copper, which was also observed by Harlander et al.\cite{Harlander:10}, is inconsistent with the photoelectric effect hypothesis, suggesting other mechanisms in effect. The charging time constant is typically shorter in copper traps, 100-200~s. The saturated micromotion amplitudes for aluminum is $\sim$15 times higher than copper at 405~nm and 5 times higher than copper at 460~nm. In the gold trap, some charging is observed at 370~nm, and not observed for any other wavelengths tested. Fitting to the cases where the micromotion signal appears to stay constant over the measurement time of 1000~s indicates a measurement sensitivity of 0.01~$\mu$m.

Comparisons of $\Delta d$ and $\tau$ for these materials and wavelengths are listed in Table \ref{Tbl:Data}. Errors are estimated from repeating the same measurements on different days with the same trap. In some cases such as aluminum at 460~nm and gold at 370~nm, the micromotion signal vs time appears closer to linear, suggesting that the time constant of charging is very long, $>$1000~s. These data are marked with (*) in Table~\ref{Tbl:Data}. After blocking the beam, the micromotion amplitude stays constant for at least 20 minutes, suggesting that discharging occurs on a much longer time scale than charging. By comparison, in previous work the discharging time constants were measured to be 654~s for aluminum\cite{Allcock:11} and 120~s for copper\cite{Harlander:10}. This is consistent with the expectation that the conductivity of the oxide material becomes negligibly small at cryogenic temperatures. Because the rate of discharge is slow, it's possible that the charges created in one experiment continue to contribute a screening effect to the subsequent measurement and thus the measurements taken on the same day are not independent of each other. To minimize such effects, we measured the wavelength dependence starting with the longest wavelength, and in the cases where only the shortest wavelengths exhibited significant charging, the screening effect should be minimal between successive measurements.

\begin{table}
\centering\begin{tabular}{| l | c | c | c | c | c | c | c | c |}
\hline
Trap & \multicolumn{2}{|c|}{370~nm} & \multicolumn{2}{|c|}{405~nm} & \multicolumn{2}{|c|}{460~nm} & \multicolumn{2}{|c|}{674~nm} \\ 
\hline
     & $\Delta d$ & $\tau$ & $\Delta d (\mu m)$ & $\tau (s)$ & $\Delta d$ & $\tau$ & $\Delta d$ & $\tau$ \\
\hline
Al        & $\circ$ & $\circ$ & 0.25(5)  & 500(70)   & *    & *       & -      & -   \\         
Al-10     & $\circ$ & $\circ$ & 0.28(11) & 770(140)  & *    & *       & -      & - \\
Al-20     & $\circ$ & $\circ$ & 0.34(2)  & 420       & *    & *       & -      & - \\
Cu        & $\circ$ & $\circ$ & 0.02     & 300(30)   & 0.05        & 80(5)   & 0.034  &  100(30) \\
Au        &  * & * &  *  & * & - & - & - & - \\
\hline  
\end{tabular}
\caption{\label{Tbl:Data} Summary of fit parameters from charging data in aluminum, copper, and gold traps. Number after ``Al'' indicate the thickness of additionally evaporated aluminum oxide. Errors are estimates based on different measurements performed on the same trap. Values without errors indicate that only one measurement was done and errors from the fit are very small, unlikely to represent actual uncertainties. Dashes (-) indicate that the fitted charging rate is consistent with zero, given that the ion displacement at the end of the measurement time is within the measurement resolution of 0.01~$\mu$m. Asterisks (*) indicate that fitting to an exponential function resulted in poor constraint on the fitting parameters, and a linear fit is used with different parameters. Circles ($\circ$) indicate that the data was not obtained.}
\end{table}

\subsection{Quantifying charge}
\label{sec:resultsB}

The charging measurements are performed by displacing the laser from directly below the ion by 25(5)~$\mu$m in order to enhance the detected micromotion signal due to ion movement. For one aluminum trap, the measurement was repeated with the laser displaced on either side of the ion. From the sign of the change in compensation voltages needed to minimize micromotion of the ion after charging, one can determine the direction of the movement of the ion and the sign of the charge. We find that the voltage on the electrode closest to the charging laser needs to be increased to re-compensate the ion, indicating that the ion moves toward the electrode and the laser beam due to charging. These observations agree with the hypothesis that the sign of the light-induced charge is negative.

\begin{figure}
\includegraphics[height=3.3in, angle=270]{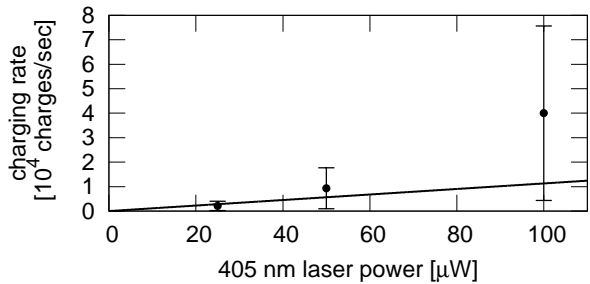}
\caption{\label{Fig:power} Charging rate $K$ (at $t=0$) measured in an aluminum trap with 20~nm of oxide on the surface, as a function of incident 405~nm laser power.}
\end{figure}

The number of charges created by the laser can be estimated by considering the trap and experiment geometry (Eqs.~1 and 2), assuming a linear relationship between the initial charging rate and the laser power. Such a relationship was observed previously\cite{Harlander:10}. Fig.~\ref{Fig:power} shows the result of such measurements, but note that the slow rate of discharging in the cryogenic environment means that measurements of charging rate vs power may not be independent. The data shown in Fig.~\ref{Fig:power} cannot conclusively rule out either a single-photon or two-photon processes for the charging effect in aluminum. Nevertheless we give an order-of-magnitude estimate for the number of charges produced from the geometry as follows. The center of the gaussian profile of charges is located $\sim$100~$\mu$m below and ~25(5)~$\mu$m to the side of the ion. The size of the dipole $r_d$ is unknown, but physical estimates of $r_d \simeq$ $1$ to $10$~nm (corresponding to twice the thickness of the oxide layer) results in a dipole density of $\sim 4 \times 10^6$ dipoles/cm$^2$ at saturation for the data of the aluminum trap at 405~nm. The initial charging rate as calculated from data fitting is $K \simeq$ 1.5$\times$10$^4$ charges/sec. The charging efficiency is then estimated to be $\eta \simeq 10^{-10}$ charges/photon. The charging efficiency for Cu and Au is not calculated since we don't have a good estimate of the dipole length.

\section{Conclusion}   

We have observed and characterized effects of laser-induced charging on microfabricated aluminum, gold, and copper ion traps. In aluminum, charging is only clearly observed for the shortest tested wavelength of 405~nm, suggesting a mechanism dominated by the photoelectric effect. No significant variation is observed for aluminum traps with varying amounts of deposited aluminum oxide. Copper traps exhibit less charging at 405~nm, but some charging is observed at all wavelengths. No charging is observed in gold traps except at 370~nm, consistent with both its higher work function compared to aluminum and copper, and the absence of a native oxide. These measurements suggest that gold may be a preferrable material for small-scale ion trap quantum computing.

In surface-electrode traps it is difficult to avoid hitting the trap surface during routine laser alignment when loading ions, but with long ion lifetimes ($\sim$ few hours in our system) and otherwise stable trapping voltages, the problems with charging may be mostly avoided. However, with lasers at shorter wavelengths (such as those needed for most species other than Sr$^+$ currently considered for trapped-ion quantum computing) or smaller ion heights, the charging issue may have greater impact. The timescales of charging observed in our experiments ($\sim$ 100s of seconds) are long compared to most gate operations ($\sim \mu$s-ms), but become relevant in experiments that require many repeated measurements over long periods of time (minutes to hours), such as precision measurements\cite{Rosenband:08} or process tomography\cite{Monz:09, Haeffner:05}. In such cases, care should be taken to detect and correct for changes in micromotion and ion position due to charging. 

We thank John Martinis and Jeffrey Russom for helpful discussions. This work was supported in part by the NSF Center for Ultracold Atoms, the IARPA COMMIT program, and the IARPA SQIP project. P.F.H. is grateful for the support from the Carlsberg Foundation and the Lundbeck Foundation.

%

\end{document}